\documentclass[useAMS,usenatbib]{mn2e}  
  
\def\e{$\pm$}  
\newcommand{\ltsima} {$\; \buildrel < \over \sim \;$} 
\newcommand{\simlt}  {\lower.5ex\hbox{\ltsima}}            
\newcommand{\gtsima} {$\; \buildrel > \over \sim \;$} 
\newcommand{\simgt}  {\lower.5ex\hbox{\gtsima}}            
 
\title[Flickering of T~CrB]{Flickering  variability of  T~Coronae Borealis   
\thanks{based on observations obtained in NAO Rozhen, Bulgaria}}  
\author[Zamanov, Bode, Stanishev, Mart\'\i]  
{R. Zamanov$^{1}$\thanks{e-mail: rz@astro.livjm.ac.uk; mfb@astro.livjm.ac.uk; vall@physto.se;  jmarti@ujaen.es},    
M. F. Bode$^{1}$, V. Stanishev$^{2}$, J. Mart\'\i$^{3}$\\  
$^{1}$ Astrophysics Research Institute, Liverpool John Moores University, Twelve Quays House, Birkenhead, CH41 1LD, UK\\  
$^{2}$ Institute of Astronomy, Bulgarian Academy of Sciences,   
       72 Tsarighradsko Shousse Blvd., 1784 Sofia, Bulgaria \\  
$^{3}$ Universidad de Ja\'en, Departamento de F{\'\i}sica (EPS),   
       Virgen de la Cabeza, 2, E-23071 Ja\'en, Spain \\  
}  
  
\begin{document}  
  
\date{Accepted . Received 2004  January 30; in original form 2003  November 11}  
  
\pagerange{\pageref{firstpage}--\pageref{lastpage}} \pubyear{2004}  
  
\maketitle  
  
\label{firstpage}  
  
\begin{abstract}  
We present electro-photometric UBV and high-speed U-band flickering 
observations of  the recurrent nova T~CrB during a period 
when its U brightness varies by more than 2 mag. The V band is dominated by the ellipsoidal  
variability of the red giant, however, the variability of  
the hot component also causes $\sim 0.15$ mag variations in V.
We define a set of parameters which characterise the flickering. 
The Fourier spectra of all 27 nights are similar to each other. The
power spectral density of the variations has a power law component 
($\propto$f$^{-1.46}$ on average). 
We do not detect a dependence of the Fourier spectra and autocorrelation 
function on the brightness of the object.  
Having subtracted the contribution of the red giant, we show that  
the flickering amplitude correlates with the average flux of the accreting  
component. A comparison with CH~Cyg and MWC~560   
indicates that the flickering of T~CrB is more stable 
(at least during the time of our observations), 
than that in the jet-ejecting symbiotic stars.  
The data are available in electronic form from the authors. 
\end{abstract}  
  
  
\section{Introduction}  
T~CrB (HD 143454) is an interacting binary star   
which consists of a red giant and a white dwarf   
(Selvelli et al. 1992,   
 and references therein).  
The star has undergone two   
nova eruptions  (Nova CrB 1866, 1946) and is thus classified  
as a recurrent nova (and, due to the presence of the cool giant, plus emission lines seen at outburst, also as a symbiotic star).  
The red giant fills Roche lobe, and thus the accretion flow onto  
the white dwarf (WD) is via L$_{1}$, which is typical for cataclysmic variables.   
Sharing characteristics of three (partly overlapping)   
types of interacting binaries, T~CrB  
is therefore an important object for our understanding of the different processes in interacting binaries.   
  
Stochastic brightness variations (flickering),   
occurring on time scales of seconds and minutes with amplitudes ranging   
from a few millimagnitudes up to more than an entire magnitude  
are a phenomenon typical for cataclysmic variables,   
and it is rarely observed in symbiotic stars. For example, to-date it is detected  
in only 8 of the  220 known symbiotics   
(Dobrzycka et al. 1996; Belczy{\'n}ski et al 2000, Sokoloski et al. 2001).  
In T~CrB flickering with amplitude of $\Delta$U$\sim$0.1-0.5 mag  
has been observed on a time scale of minutes   
(Ianna 1964, Lawrence et al. 1967  
Bianchini \& Middleditch 1976, Walker 1977,  
Bruch 1980). The flickering amplitude is somewhat smaller   
in B and V bands (Raikova \& Antov 1986, Hric et al. 1998).   
 In addition, on some occasions such flickering disappears   
(Bianchini \& Middleditch 1976, Oscanian 1983, Miko{\l}ajewski et al. 1997).   
In our previous investigation (Zamanov \& Bruch 1998) we showed  
that the flickering of T~CrB is indistinguishable from the flickering  
observed in dwarf novae, in spite of the vast difference   
in the  geometrical size of the systems.  
  
 The exact origin of the stochastic variations is not clear, but they are  
considered to be a result of accretion onto the WD through a disk. The possible mechanisms   
include unstable mass transfer, magnetic discharges, turbulence and instability   
in the boundary layer (e.g. Warner 1995, Bruch 1992).  

Here we present new UBV and high-speed flickering observations of T~CrB,  
estimate the contribution of the red giant,  
analyse the U band variability, search for relations between   
the flickering quantities and the brightness of the object,  
and compare its behaviour with two other symbiotic stars   
(the "nanoquasars" CH~Cyg and MWC~560).   
  
\section{Observations}   
  
 \begin{table*}  
 \caption{UBV observations of T~CrB. }  
  \begin{tabular}{@{}cccccccccccccc@{}}  
\hline
 JD-2400000  &    V   &   B     &     U   &  & JD-2400000  &	V   &	B     &     U	 & & JD-2400000  &    V   &   B     &	  U	  \\  
\hline
  50476.655  & 10.309 &  11.570 & 11.568  &  &	 50698.295  & 10.146 &  11.298 & 11.089  & &  51008.314  & 10.062 &  11.393 & 11.681	    \\  
  50476.659  & 10.311 &  11.564 & 11.620  &  &	 50698.301  & 10.142 &  11.269 & 10.979  & &  51009.304  & 10.026 &  11.318 & 11.574	    \\  
  50477.556  & 10.299 &  11.476 & 11.367  &  &	 50739.236  &  9.946 &  10.960 & 10.655  & &  51009.308  & 10.072 &  11.403 & 11.720	    \\  
  50477.559  & 10.280 &  11.507 & 11.496  &  &	 50739.242  &  9.943 &  11.075 & 11.105  & &  51015.323  & 10.055 &  11.343 & 11.526	    \\  
  50478.580  & 10.276 &  11.490 & 11.363  &  &	 50741.222  &  9.773 &  10.777 & 10.349  & &  51015.327  & 10.073 &  11.353 & 11.442	    \\  
  50478.583  & 10.274 &  11.446 & 11.314  &  &	 50741.227  & 10.036 &  11.950 & 12.671  & &  51016.308  & 10.080 &  11.387 & 11.566	    \\  
  50479.660  & 10.315 &  11.558 & 11.515  &  &	 50828.591  & 10.213 &  11.245 & 11.087  & &  51016.313  & 10.086 &  11.395 & 11.643	    \\  
  50479.663  & 10.311 &  11.579 & 11.517  &  &	 50828.595  & 10.188 &  11.217 & 11.144  & &  51027.320  & 10.124 &  11.412 & 11.531	    \\  
  50480.651  & 10.322 &  11.693 & 11.780  &  &	 50864.466  & 10.104 &  11.346 & 11.600  & &  51027.324  & 10.160 &  11.413 & 11.497	    \\  
  50480.654  & 10.370 &  11.615 & 11.639  &  &	 50864.470  &  9.852 &  11.226 & 11.508  & &  51034.298  & 10.215 &  11.468 & 11.602	    \\  
  50504.566  & 10.198 &  11.558 & 11.850  &  &	 50865.491  & 10.015 &  11.294 & 11.476  & &  51034.302  & 10.231 &  11.470 & 11.583	    \\  
  50504.572  & 10.162 &  11.425 & 11.676  &  &	 50865.499  &  9.993 &  11.280 & 11.406  & &  51226.574  &  9.848 &  11.063 & 11.085	    \\  
  50520.518  & 10.025 &  11.456 & 12.039  &  &	 50867.519  &  9.974 &  11.322 & 11.550  & &  51226.582  &  9.911 &  11.167 & 11.328	    \\  
  50520.524  & 10.040 &  11.433 & 11.970  &  &	 50867.523  &  9.958 &  11.313 & 11.484  & &  51239.592  & 10.041 &  11.356 & 11.555	    \\  
  50628.448  &  9.982 &  11.084 & 10.809  &  &	 50877.603  &  9.974 &  11.373 & 11.658  & &  51239.597  & 10.012 &  11.303 & 11.387	    \\  
  50628.453  &  9.972 &  11.048 & 10.766  &  &	 50877.609  & 10.023 &  11.387 & 11.660  & &  51401.292  & 10.054 &  11.095 & 10.714	    \\  
  50651.337  &  9.911 &  11.184 & 11.319  &  &	 50877.620  & 10.023 &  11.414 & 11.773  & &  51401.297  & 10.019 &  11.040 & 10.637	    \\  
  50651.342  &  9.906 &  11.191 & 11.328  &  &	 51005.484  &  9.841 &  10.828 & 11.412  & &  51404.314  & 10.224 &  11.539 & 11.637	    \\  
  50652.325  &  9.820 &  10.919 & 10.714  &  &	 51005.485  &  9.853 &  10.833 & 11.371  & &  51404.318  & 10.274 &  11.649 & 11.787	    \\  
  50652.329  &  9.785 &  10.919 & 10.730  &  &	 51007.487  & 10.095 &  11.390 & 11.691  & &  51408.289  & 10.249 &  11.712 & 12.123	    \\  
  50654.307  &  9.844 &  10.948 & 10.766  &  &	 51007.493  & 10.074 &  11.415 & 12.143  & &  51408.294  & 10.267 &  11.689 & 12.070	    \\  
  50654.311  &  9.733 &  10.908 & 10.811  &  &	 51008.310  & 10.090 &  11.440 & 11.715  & &	 	&	 &	   &		   \\  
\hline
  \label{tUBV}									   				      
  \end{tabular}									   				      
  \end{table*}	  


The observations have been performed  with the 60 cm   
telescope of NAO Rozhen equipped with a single channel photometer.   
The comparison stars  
were HD142929 and BD$+26{^0}2761$, the check star GSC 2037.1228  
and the integration time 1 or $10\ sec$. The observations   
with 1 sec integration time have been binned in 10 sec.   
APR software (Kirov, Antov \& Genkov, 1991) has been used for   
data processing.   
The accuracy of the UBV photometry is better than 0.03 mag and the results  
are given in Table~\ref{tUBV}.  
  
For the flickering observations,  
the reduction to the standard U band is better  
than $\pm 0.04\,$ mag and the  internal  accuracy  of  the  data  
(standard deviation from the average of 10 consecutive measurements)   
is  $0.015-0.030\,$ mag.  
The control of the atmospheric conditions and performance of the system
have been done by observing the check star, before and after T CrB.
and  carefully tracing of the  comparison star counts 
(which has been observed every ~20-30 minutes).
In the subsequent data processing  2 nights has been rejected, because of
"doubtful" behaviour of the comparison or/and check stars. 
Journal of flickering observations and the main characteristics   
of the U band variability for each run are summarised in Table~\ref{jour}.  
 
The error in the magnitudes (U$_{max}$, U$_{min}$, U$_{av}$)  
are calculated dividing every run into two parts and calculating the   
quantities separately for each part, in this way addressing the possible   
errors of the run.   
In Fig.\ref{figV} is plotted the orbital modulation in V, 
in Fig.\ref{fl1} the long term U band curve and the flickering observations, 
and in Fig.\ref{exam} are given two examples of the flickering.
 
\section{Contribution of the red giant}  

 \begin{centering}  
 \begin{table*}  
 \caption{Journal of flickering observations in the U band.   
 Date is given in format YYMMDD, TJD is the truncated Julian day  
 of the start of the observation, N is the number of the points in the run, IT is the integration time [in seconds].  
 D is the duration of the run in minutes.  U$_{max}$, U$_{min}$, and U$_{aver}$  
 are the maximum brightness during the night, the minimum, and the average of this quantity, respectively. 
 $\sigma$ is the standard deviation. 
 U$_{av}$ is calculated averaging the corresponding fluxes. The power spectra in each night 
 are fitted with linear fit (A and $\gamma$ are the parameters of the fit, see the text). 
 $\gamma$ is the power spectrum slope  in the interval 3-160 cycles/hour.    
 The e-folding time of the ACF is given for the original run ($\tau_0$), and after   
 subtraction of a spline fit ($\tau_1$).  
 }  
  
\begin{tabular}{rlrrrcccccrr}  
\hline
 Date   & TJD$_{start}$ &  NxIT    &  D        & U$_{max}$  & U$_{min}$ & U$_{av}$ &  $\sigma$ & A	& $\gamma$	&  $\tau_0$ & $\tau_1$     \\  
        &               &  [sec]   &  min      & [mag]      & [mag]     &  [mag]   &  [mag]    &	&   &  [sec]	&   [sec]      \\  
\hline
930228  &   49046.5117  &  448x10 & 122  & 12.260\e0.03  & 12.578\e0.01   &  12.387\e0.04 &   0.084 &  3.41& $-$1.77 &  789$\pm$35   &   65$\pm$7    \\  
940410  &   49453.4621  &  266x10 &  49  & 12.652\e0.02  & 12.945\e0.05   &  12.820\e0.05 &   0.081 &  3.59& $-$1.71 &  516$\pm$72   &   54$\pm$10   \\  
950613  &   49882.3317  &  242x10 &  45  & 11.713\e0.11  & 12.222\e0.07   &  11.952\e0.08 &   0.121 &  3.19& $-$1.32 &  257$\pm$49   &  143$\pm$21   \\  
950620  &   49889.3496  &  520x10 & 102  & 11.798\e0.06  & 12.353\e0.10   &  12.081\e0.05 &   0.099 &  2.14& $-$1.47 &  181$\pm$25   &  107$\pm$10   \\  
960110  &   50092.6671  &  150x10 &  31  & 11.826\e0.07  & 12.236\e0.03   &  12.002\e0.03 &   0.092 &  3.70& $-$1.38 &  138$\pm$87   &   98$\pm$17   \\  
960228  &   50141.5088  &  772x10 & 162  & 11.161\e0.02  & 11.714\e0.01   &  11.406\e0.02 &   0.121 &  2.23& $-$1.51 &  292$\pm$18   &  130$\pm$36   \\  
960229  &   50142.4900  &  9304x1 & 190  & 11.460\e0.01  & 12.065\e0.07   &  11.770\e0.05 &   0.105 &  1.74& $-$1.40 &  1253$\pm$82   &  131$\pm$12   \\  
960325  &   50167.5358  &  3956x1 &  63  & 12.017\e0.01  & 12.323\e0.02   &  12.146\e0.00 &   0.052 &  3.04& $-$1.68 &  132$\pm$30   &  105$\pm$21   \\  
961216  &   50433.6246  &  2190x1 &  57  & 11.051\e0.00  & 11.397\e0.03   &  11.272\e0.02 &   0.078 &  1.52& $-$1.06 &  152$\pm$21   &  132$\pm$18   \\  
961218  &   50435.6354  &  1292x1 &  38  & 11.111\e0.04  & 11.368\e0.03   &  11.262\e0.04 &   0.066 &  2.50& $-$1.41 &   66$\pm$6    &   78$\pm$7    \\  
970128  &   50476.5542  &  674x10 & 128  & 11.181\e0.05  & 11.632\e0.10   &  11.456\e0.07 &   0.101 &  2.70& $-$1.45 &  675$\pm$89   &  174$\pm$10   \\  
970129  &   50477.5717  &  522x10 &  98  & 11.307\e0.01  & 11.607\e0.02   &  11.469\e0.02 &   0.070 &  2.56& $-$1.47 &  378$\pm$47   &  144$\pm$22   \\  
970130  &   50478.5983  &  425x10 &  81  & 11.333\e0.07  & 11.614\e0.04   &  11.461\e0.05 &   0.068 &  2.98& $-$1.54 &  785$\pm$204  &   50$\pm$8    \\  
970131  &   50479.5537  &  704x10 & 137  & 11.354\e0.05  & 11.631\e0.04   &  11.488\e0.04 &   0.060 &  2.95& $-$1.67 &  416$\pm$25   &  185$\pm$13   \\  
970201  &   50480.5379  &  8318x1 & 151  & 11.443\e0.01  & 11.828\e0.02   &  11.658\e0.04 &   0.093 &  2.42& $-$1.54 &  626$\pm$61   &  169$\pm$17   \\  
970721  &   50651.3508  &  547x10 & 119  & 11.200\e0.02  & 11.457\e0.01   &  11.326\e0.01 &   0.064 &  2.52& $-$1.43 &  799$\pm$38   &   50$\pm$5    \\  
970722  &   50652.3379  &   96x10 &  21  & 10.533\e0.01  & 10.762\e0.03   &  10.669\e0.03 &   0.068 &  3.80& $-$1.73 &  114$\pm$13   &  135$\pm$16   \\  
970827  &   50688.3313  &  207x10 &  46  & 10.561\e0.06  & 10.939\e0.07   &  10.758\e0.06 &   0.086 &  3.57& $-$1.67 &  280$\pm$60   &  123$\pm$22   \\  
980220  &   50864.5092  &  837x10 & 168  & 11.464\e0.01  & 11.779\e0.02   &  11.604\e0.01 &   0.071 &  2.14& $-$1.40 &  571$\pm$44   &  124$\pm$12   \\  
980224  &   50868.5383  &  407x10 &  81  & 11.296\e0.06  & 11.700\e0.07   &  11.505\e0.05 &   0.094 &  3.11& $-$1.58 &  498$\pm$45   &  139$\pm$23   \\  
980713  &   51008.3217  &  256x10 &  52  & 11.540\e0.01  & 11.774\e0.00   &  11.660\e0.00 &   0.062 &  2.76& $-$1.28 &  212$\pm$25   &  256$\pm$46   \\  
980714  &   51009.3188  &  362x10 &  74  & 11.504\e0.04  & 11.840\e0.02   &  11.670\e0.05 &   0.078 &  2.94& $-$1.43 &  257$\pm$216  &   89$\pm$10   \\  
980720  &   51015.3342  &  304x10 &  64  & 11.335\e0.02  & 11.621\e0.00   &  11.477\e0.01 &   0.072 &  3.34& $-$1.67 &  191$\pm$14   &  125$\pm$18   \\  
980721  &   51016.3208  &  319x10 &  64  & 11.435\e0.03  & 11.770\e0.05   &  11.613\e0.02 &   0.059 &  2.72& $-$1.41 &  105$\pm$14   &   91$\pm$11   \\  
980802  &   51028.3125  &  344x10 &  69  & 11.400\e0.06  & 11.759\e0.02   &  11.595\e0.04 &   0.079 &  2.19& $-$1.09 &  387$\pm$118  &  106$\pm$14   \\  
980803  &   51029.3217  &  298x10 &  61  & 11.521\e0.02  & 11.737\e0.02   &  11.614\e0.01 &   0.042 &  2.81& $-$1.38 &   87$\pm$25   &   40$\pm$10   \\  
990107  &   51185.6225  &  406x10 &  76  & 11.646\e0.07  & 12.074\e0.05   &  11.847\e0.05 &   0.086 &  2.33& $-$1.40 &  495$\pm$144  &   54$\pm$5    \\  
\hline
\label{jour}									   				      
\end{tabular}									   				      
\end{table*}	  
\end{centering}   

\subsection{V band}  
\label{VgM}  
In symbiotic stars, the mass donor is a red giant. In the case of T~CrB  
its contribution is not negligible in the UBV bands.   
The V band variability of T~CrB is dominated by the ellipsoidal variability  
of the red giant (Peel 1985, Lines et al 1988).   
The V band data from the long term light curve   
(see Stanishev et al. 2004 and references therein)   
are plotted in Fig.1, folded with the orbital period.   

A three term truncated Fourier fit to all data gives   
\begin{eqnarray*}  
& & V =  10.056\,(0.003) \\
&  & +0.007\,(0.004)\cos 2\pi\phi  -0.026\,(0.004)\sin 2\pi\phi  \\  
&  & -0.161\,(0.004)\cos 4\pi\phi  -0.036\,(0.004)\sin 4\pi\phi   \\
&  & +0.016\,(0.004)\cos 6\pi\phi  -0.037\,(0.004)\sin 6\pi\phi 
\end{eqnarray*}  
where $\phi$ is the orbital phase (hereafter the numbers in the parentheses
refer to the errors).  This fit is plotted in Fig.1.  
Typical deviation of the points from the fit line is $\pm$0.10 mag.  
  
\begin{figure}  
 \mbox{}  
 \vspace{6.5cm}  
  \includegraphics{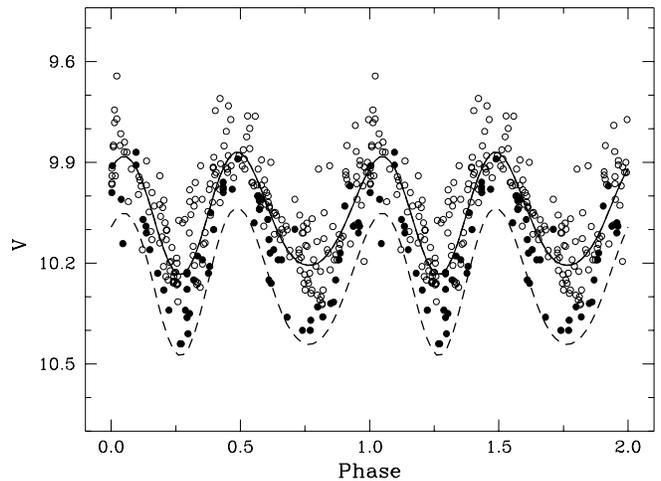}   
  \caption[]{Johnson V band observations of T~CrB  folded with a 227 day period.  
  The solid line is our fit to all data. The open circles refer to the epochs when  
  the object is brighter than U$=12$ and the filled circles refer to   
  the epochs when it is fainter than this.    
  The dashed line is the calculated contribution of the red giant.   
  }		    
\label{figV}     
\end{figure}

\begin{figure}
 \mbox{}  
 \vspace{6.0cm}  
  \includegraphics{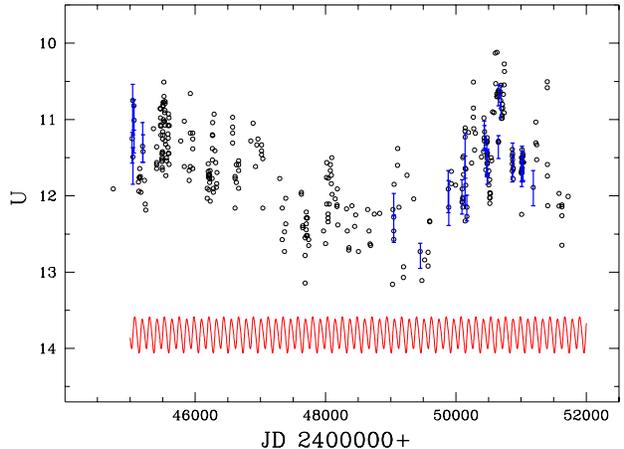}   
  \caption[]{Long term variability in U band magnitudes and the flickering observations, with the  
  corresponding amplitude. The cosine wave (bottom) is the calculated contribution of the red giant.   
  }		    
\label{fl1}     
\end{figure}

On Fig.1 we have plotted with different symbols the points when the object is   
brighter and fainter at shorter wavelengths (open circles refer to U$<$12  
and filled -- to U$\ge$12). It is visible that the filled circles are   
displaced downward slightly relative to the open ones.  
The U band brightness is dominated by the hot component. We can also deduce   
that  the variability of the hot component of about 2 magnitudes in U   
(see also Stanishev et al. 2004) also contributes to that in V.   
To define this contribution  
we performed a simple fits (using only the main terms) to the open and filled symbols.  
The obtained coefficients are:  
\begin{eqnarray*}
V=10.157\,(0.005)-0.194\,(0.007)\cos 4\pi\phi  & {\rm for}  &  U\ge12   \\
V=10.029\,(0.003)-0.163\,(0.004)\cos 4\pi\phi  & {\rm for} & U < 12 .  
\end{eqnarray*}  
The mean values of the U band magnitudes are U=12.3$\pm$0.3 and U=11.2$\pm$0.4   
for the fainter and brighter points respectively. 
Therefore, the increase of the system U-band brightness    
brightness by 1.1 mag results in an increase of the V brightness by
0.128 (Eq. 3,4).
We derive a relative contribution   
R(V)=0.205$\pm$0.035, where R(V)$=$F$_{hot}$/F$_{gM}$ is the relative   
contribution between the accreting object and the red giant at V=10.056.   
The corresponding orbital light curve of the red giant is plotted   
as a dashed line on Fig.\ref{figV}.   
The calculated contribution is very similar to that  obtained by Zamanov \& Bruch (1998)   
on the basis of the average colours of the flickering source in cataclysmic variables.   
  
Although the data in Fig.\ref{figV}  spread over 22 years, the   
typical deviation of the points from the fit line is $\pm$0.10 mag.  
This points to the fact that the $V$ band light curve  
has not changed in its main features over the last 22 years.  
This in turn indicates that the M giant is not variable. Indeed, we can put an  
upper limit on its possible variability of $\Delta V <$ 0.05 -- 0.10.   
The stability of the red giant is better defined  in IR   
observations (Yudin \& Munari 1993, Shahbaz et al., 1997), 
where a upper limit of variability  $\Delta J <0.02$ has been constrained. 
  
\subsection{Red giant contribution to U band flux}  
  
The latest definitions of the spectral type of the red giant   
in T~CrB are  M4III (Zhu et al. 1999) and   
M4.5III (M{\" u}rset \& Schmid 1999). Both are obtained on the basis of   
IR spectra and with typical uncertainty $\pm 1$ spectral subtype.   
The expected colour of a  M4.5III star is 
(U-V)$_{M4.5III}=3.16 \pm 0.10$ (Lee 1970), 
(U-V)$_{M4.5III}=3.28 \pm 0.10$ (Schmidt-Kaler  1982), 
(U-V)$_{M4.5III}=3.25 \pm 0.10$ (calculated using the tables of Fluks et al. 1994). 
The New ATLAS9 model atmospheres 
(Pietrinferni, Cassisi, Salaris, Castelli, in preparation), 
for a star T$_{eff}=$3500~K and $\log (g) = 1.0$, give 
(U-V)=3.18 for [Fe/H]=0.2,
(U-V)=3.26 for [Fe/H]=0, and 
(U-V)=3.42 for [Fe/H]=-1.5.
  
Using R(V)=0.222 at V=10.029 (as derived in Section \ref{VgM}), 
the fit to V (Eq.1), E$_{B-V}=0.15$,  and  (U-V)$_{M4.5III} =$3.25,  
we can calculate the contribution of the red giant   
to the U band. The light curve of T~CrB, our flickering observations,   
and the red giant contribution (as a sine wave) are plotted in Fig.2.  
  

One way to check the calculated U brightness of the red giant 
is IR photometry, supposing that the red giant is the only source in J band.
The J brightness of T~CrB varies in the interval J=5.90 -- 6.19 
(Kamath and Ashok 1999). 
Interpolations in the tables of the colours of red giants  give 
(U-J)$_{M4.5III}=7.75 \pm 0.25$ (from Lee 1970), 
(U-J)$_{M4.5III}=7.14 \pm 0.2$  combining  (U-V) from  
Schmidt-Kaler (1982)  and (V-J) from Ducati et al. (2001),
(U-J)$_{M4.5III}=7.34 \pm 0.2$ (from  Fluks et al. 1994), 
where the calculated uncertainties refer 
to $\pm 0.5$ spectral type.
The model atmospheres give (U-J)$_{[Fe/H]=0.2}$=7.24, 
(U-J)$_{[Fe/H]=0}$=7.13 
and (U-J)$_{[Fe/H]=-1.5}$=6.52 (Pietrinferni et al. 2004).

Shahbaz et al (1999) modeled the M giant spectrum,  
with enhanced  abundance of lithium but normal abundance of the other metals,  
however they used  higher gravity inconsistent 
with the last orbital solution.

If we suppose that the red giant is the only source in J, 
using the above (U-J) colours and  
E$_{B-V}=0.15$ we could expect  value 
about  U$\sim$ 14.4 - 13.3, which is
consistent with the supposed contribution of the red giant  
to the U band (see Fig.2).

  
\section{Flickering quantities}  
 \begin{figure}
 \mbox{}  
 \vspace{5.0cm}  
  \includegraphics{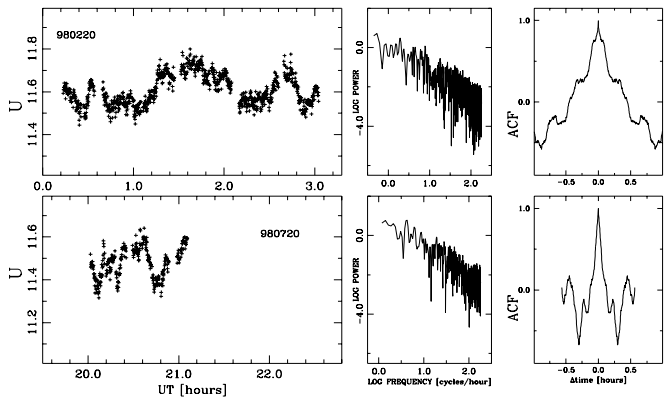}   
  \caption[]{Two examples of our observations (for dates 980220 and 980720).   
  The left panels represent the flickering behaviour of T~CrB in U,   
  mid panels the power spectra, and right panels the autocorrelation function.  
  }		    
\label{exam}     
\end{figure}	    
  
  
 \begin{figure}  
 \mbox{}  
 \vspace{8.0cm}  
  \includegraphics{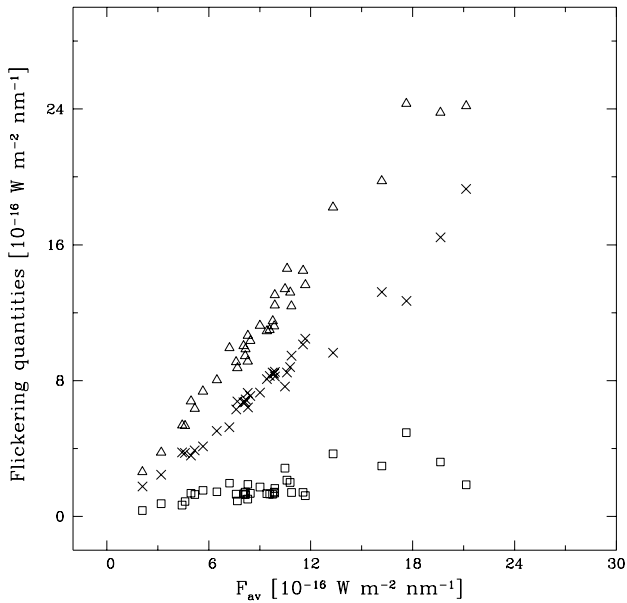}   
  \caption[]{Measured flickering quantities versus the average flux F$_{av}$   
  of the hot component.  Triangles refer to F$_{max}$, crosses to F$_{min}$ and squares to    
  F$_{fl}$. The axes are in units $10^{-16}$ ${\rm Watt\: m}^{-2}{\rm nm}^{-1}$.  
  }		    
\label{fl2}     
\end{figure}	    
  
  
The U magnitudes were converted into fluxes, adopting   
flux for a zero magnitude star   
F$_{0}(U)=4.194\times10^{-11}$ ${\rm Watt\: m}^{-2}{\rm nm}^{-1}$  
(Bessel 1979).   
In addition to our data, we used those from Bruch (1992)  
and the data of Oskanian (1983). Bruch's data are reddened with  
$E_{B-V}=0.12$ and they were corrected for the difference in the adopted zero-point  
of the flux scale.  
For Oskanian's data, we assumed that the instrumental difference  
$\Delta u=0$ corresponds to magnitude U=11.83.  
We have used only positive detections of the flickering.  
After the observed flux during a given night was corrected   
for the contribution of the red giant the following quantities 
were calculated:
  
F$_{av}$ -- the average flux of the hot component;
  
F$_{max}$ -- the maximum flux of the hot component;
  
F$_{min}$ -- the minimum flux of the hot component;  
  
F$_{fl}$ -- the average flux of the flickering, F$_{fl}=$F$_{av} -$F$_{min}$.

In Fig.\ref{fl2} are plotted the flickering quantities versus  the average flux of the  
hot component. It is obvious they have to be connected,   
however it is not clear a priori how   
the different quantities will depend on each other.  
The least squares fits to data in Fig.4  give:  
\begin{eqnarray}  
F_{max}= -0.22\,(0.07) + 1.252\,(0.006)\,\,F_{av} \\  
F_{min}= -0.32\,(0.56) + 0.858\,(0.054)\,\,F_{av} \\  
F_{fl} = +0.19\,(0.42) + 0.156\,(0.042)\,\,F_{av}   
\end{eqnarray}

\subsection{Power spectra}  
 
For each run we also calculated the power spectrum and the autocorrelation  
function. Two examples are shown in Fig.\ref{exam}.  
Over a wide range of frequencies the power spectra of T~CrB light curves
follow power law $P(f)\propto f^\gamma$,  where $P$ is the power and 
$f$ is the frequency. Such a power-law shape is 
commonly observed in the light curves of cataclysmic variables and is attributed
the flickering.
The power-law index $\gamma$ was determined 
in the frequency interval from 3 to 160 cycles/hour.
Practically, we fitted the  
power spectra over this interval in log-log scale  
with least-squares linear fit: $\log(P)=A + \gamma \log(f)$.
$\gamma$ and A are given in Table 2. 
The typical errors are $\Delta$A$=\pm 0.8$ and  $\Delta \gamma=\pm 0.25$.
The visual comparison/inspection  
shows that all power spectra are very similar.   
This is confirmed from the fits. They give an  
average value of $\overline{\gamma}= -1.46 \pm 0.17$.    
There are two runs where $\gamma$ is $\approx -$1.1.   
In both cases  
two stronger flashes are visible in the variability with amplitude about $\sim$ 0.2 mag.  

The power-law spectrum of the type observed in T CrB 
is expected in the model of flickering proposed by 
Yonehara, Mineshige, \& Welsh(1997). They proposed as origin of flickering 
self-organised critical state of the disk in which seemingly chaotic fluctuations 
can be produced. Such a model implies $\gamma \simeq -1$ -- $-2$.
Our observations of T~CrB do not contradict to this model. 
 
The correlation analysis shows no correlation of A and $\gamma$  with  U$_{av}$ or F$_{av}$,  
(r$_P < 0.2$), indicating that the flickering is "stable", i.e. without considerable   
changes of the nature of the power spectrum in spite of the variability.


This is however not the situation in the symbiotic star CH~Cyg, where the   
power spectrum changes dramatically. There are even moments  
when CH~Cyg's power spectrum cannot be fitted with a simple power-law  
as a result of an unstable disk and disruption of the inner disk during the jet ejection 
(Sokoloski \& Kenyon, 2003b).  
In T~CrB  we did not observe instabilities like those observed in CH~Cyg.

\subsection{Autocorrelation function (ACF)}  
  
Another objective way to investigate flickering behaviour is   
the ACF (see also Bruch 2000). The ACFs were calculated according to   
Edelson \& Krolik (1988) for unevenly spaced data.  
  
The typical time scale of the flickering  may be defined as the time   
shift at which the ACF  first reaches the value  1/e.   
Thus determined correlation times are influenced by the presence   
of periodic brightness variations or some trends in the data.   
As Robinson \& Nather (1971) and Panek (1980) note, these correlation times   
can be additionally biased by the presence of weakly correlated noise   
and the process of trend removal (if applied).   
  
The e-folding time is given in Table 2.  
The errors are determined from the errors of the autocorrelation coefficients.   
We calculated e-folding times in two different ways:   
(i) the e-folding time of the ACF of the original data in each run ($\tau_{0})\;$  
and  (ii) after subtraction of a spline fit $(\tau_{1})$.  
Operation (ii) has been done in order to  obtain the typical time of the flickering   
on shorter time scales. A tension spline interpolation was undertaken. 
We subtracted this spline fit through the mean points in non-overlapping bins 
of length about  $\sim$20 minutes in a way that  
is identical to that applied to TT Ari by Kraicheva et al. (1999).   
  
The e-folding time of the ACF varies over a wide interval.             
The average values, standard deviation of the average,   
and median value of the e-folding time are        
$\overline{\tau_{0}}= 394 \pm 287$, $<\tau_{0}>=292$,  
$\overline{\tau_{1}}= 115 \pm  48$, $<\tau_{1}>=123$,  
where all values are in seconds.    
The values of $\tau_1$ are more or less similar to the values   
of TT Ari as defined in Kraicheva et al. (1999).  
  
The correlation analysis showed that there are no correlations between the  
so-defined e-folding times and the brightness of the object,   
or the flickering quantities (F$_{av}$, F$_{fl}$, or the size of the boundary layer).   
Linear Pearson correlation coefficient and  Spearman's Rank correlation   
give values  between  0 and 0.2,   
indicating that there is no dependence of the time scale of variability   
on the luminosity of the hot component, i.e. the   
the characteristic time scales of the flickering   
are not connected with the brightness of the object.  
  
\subsection{Boundary layer}  
\label{BL}  
The origin of flickering is still not clear although  
this phenomenon is observed for many stars.   
Bruch (1992) and Bruch \& Duschl (1993) identify the boundary layer   
between the accretion disk and the white dwarf as the most probable place for the origin  
of flickering. Bruch \& Duschl (1993) consider that the ratio   
$\frac{F_{fl}}{F_{min}}$ is connected with the size  
of the boundary layer between the white dwarf and the accretion disk.  
  
\begin{figure}  
 \mbox{}  
 \vspace{6.0cm}  
  \includegraphics{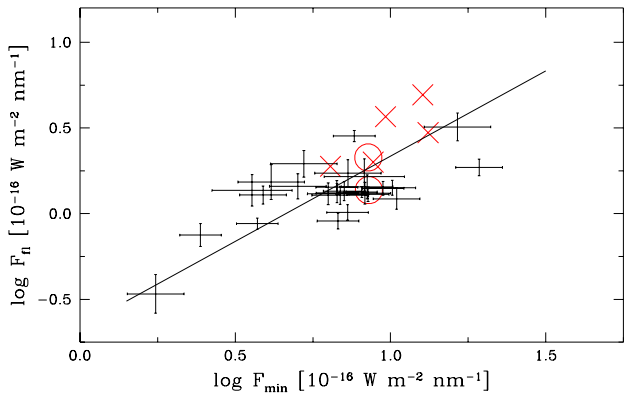}   
  \caption[]{   
  The average flux of the flickering, F$_{fl}$, versus the  quiescent    
  flux of the hot component in a light curve, F$_{min}$.  
  The solid line is a linear least squares fit $F_{fl} \propto F_{min}^k$,   
  where $k=0.995\pm 0.015$. Our data are plotted with the corresponding errors, 
  the circles represent data of Bruch (1992) and  
  crosses($\times$) - from Oskanian (1982) 
  }		    
\label{bl}     
\end{figure}

In T~CrB F$_{fl}$ is well correlated with F$_{min}$ (see Fig.\ref{bl}) . The   
Pearson's Correlation Coefficient is r$_P$=0.72 and   
Spearman's Rank Correlation r$_S$=0.56.   
Assuming that the deviation of the points from the fit lines  
(Fig 2. and Eqs. 3,4,5) is due only to the errors of the measurements,  
the fits (Eqs. 3,4,5) indicate that in spite of variations in $F_{av}$  
(which we suppose is related to the mass accretion rate),  
the ratios $ F_{fl}/F_{min}$ and  ${F_{max}}/{F_{min}}$ do not change  
markedly. In terms of Bruch \& Duschl (1993) this means that  
the size of the boundary layer remains almost constant independently of  
changes in the mass accretion rate.   
Here, adding more data (see Fig.\ref{bl}) we confirm the conclusion   
of Zamanov \& Bruch (1998) that F$_{fl}$ increases linearly with   
the increase of F$_{min}$.   
This, within the limits of Bruch \& Duschl's model,  
means that the size of the boundary layer   
in  T~CrB remains almost constant independently   
of the changes in the mass accretion rate.  
  
\section{Flickering amplitude}  
  
\begin{figure}  
 \mbox{}  
 \vspace{10.0cm}  
  \includegraphics{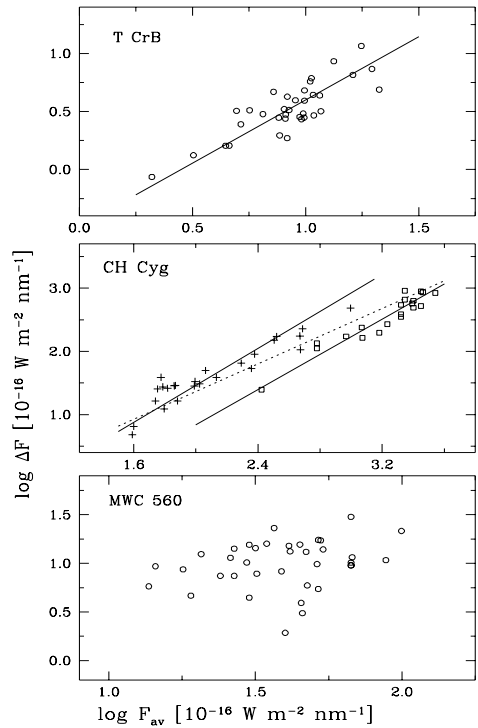}   
  \caption[]{ The flickering amplitude versus the flux of the  
      hot component in U the band for T~CrB (our data), CH~Cyg (Miko{\l}ajewski et al. 1990,  
      the plus signs refer to the propeller, the squares to the accretor stage),  
      and MWC~560 (Tomov et al. 1995).   
      The axes are in units of $10^{-16}$ ${\rm Watt\: m}^{-2}{\rm nm}^{-1}$.   
      In all cases the contribution of the red giant has been removed.  
      For T~CrB the line is the best linear fit. For CH~Cyg the solid lines   
      are for propeller and accretor stages, and the dashed line that to all points.  
      For MWC~560 the correlation is weak and no fit was performed.  
      }		    
\label{3stars}     
\end{figure}

Flickering amplitude ($\Delta F=F_{max}-F_{min}$) is also measured in the extensive observations   
of  CH~Cyg (Miko{\l}ajewski et al., 1990) and MWC~560 (Tomov et al. 1996).  
The flickering amplitude versus the average flux of the hot component, after subtraction  
of the red giant contribution  is presented  for all three stars in Fig.\ref{3stars}.

\subsection{T~CrB}  
  
The data in Fig.\ref{3stars} show that the flickering amplitude  
depends on the average hot component flux. The correlation is well defined with $r_P=0.72$ and $r_S=0.56$.  
Searching for a dependence of the type   
$\Delta F \propto (F_h)^k$, we obtain a best fit   
for T~CrB  $k=1.09\pm0.11$.  
The fit and the corresponding error have been calculated in two ways:  
(1) Using the errors into  corresponding quantities as given in Table 2. 
(2) Bootstrapping simulations (e.g. Efron \& Tibshirani, 1993)   
over the points plotted in Fig.5, i.e. taking $\sim$20 subsamples from our points.  
Using only our own and Bruch's points (i.e. data well calibrated in U) we   
obtain  $k=1.03\pm0.09$. Using different subsamples of the whole sample  
we obtained values $0.93\le k \le 1.22$. An error in the subtraction of   
the M giant contribution of 25\% will cause error in k of about 0.05.  
  
\subsection{CH~Cyg}  
\label{amplCH}  
Dependence of the flickering amplitude on the   
brightness has been reported for CH~Cyg by Miko{\l}ajewski et al. (1990).   
Their results show that the flickering amplitude in CH~Cyg  
is a power law function of the hot component luminosity,  
i.e.  $\Delta F \propto (F_h)^k$, where $k=1.40-1.45$.  
Here, using their data, we subtracted the contribution of the red giant and 
the resulting points  
are plotted  on Fig.\ref{3stars}.   
To subtract the contribution of the red giant we assume that at   
the minimum of V flux all the light is due only to the red giant, and it has   
a colour corresponding to (U-V)$_{M6III}$= 2.43 - 2.70  
(Lee, 1980; Schmidt-Kaler 1982; Fluks 1994).  
The minimum brightness of CH~Cyg is V=10.0 (Miko{\l}ajewski et al. 1996),   
and we adopt a contribution of the red giant of U=12.45.    
We will not go into details as to whether the system is triple  
(Hinkle et al. 1993; Skopal et al. 1998)   
or binary (Munari et al. 1996).   
Here we only note that an error of $\pm$0.7 mag in the subtraction of the red star(s)   
flux  would influence the obtained slope to less than $\pm$ 0.02.  
  
The analysis gives a very high correlation between   
F$_{av}$  and $\Delta$F.   
r$_P$ and  r$_S$ are always about 0.88-0.94  using (1) all points,   
(2) the propeller state observations, and (3) the accretor state   
(for further discussion of propeller and accretor states in   
CH~Cyg, see Miko{\l}ajewski et al 1990).  
  
After the subtraction of the red giant contribution   
we obtain  $k=1.08\pm0.05$ if we use all points,   
$k=1.48\pm0.05$ for propeller state points only,  
and $k=1.41\pm0.05$ for the accretor state only.   
It has to be noted that the fit $\Delta F \propto (F)^{1.08}$  
obtained on the basis of all points is very similar   
to the value for T~CrB.   
  
\subsection{MWC~560}  
  
For MWC~560 the minimum brightness is V=10.2 (Tomov et al. 1996) and  
the red giant is classified as M5.5 III (Schmid et al 2001).  
Using  (U-V)$_{M5.5III}$=2.80 - 2.95 (Lee, 1980; Schmidt-Kaler 1982; Fluks 1994),   
we  adopt a contribution of the red giant equivalent to U$\approx 13.0$.   
It deserves noting that the calculated red giant contribution   
in MWC~560 and CH~Cyg is considerably smaller than that in T~CrB,  
which is in accordance with the fact that in these two objects  
the  V  band variability is dominated by the hot component  
(not by the red giant as in T~CrB).  
  
The flickering amplitude versus F$_{av}$ for MWC~560  
is plotted in Fig.\ref{3stars} (lower panel).   
The correlation is not very significant  
(r$_P=0.25$ r$_S=0.27$ --  
although, if we delete 3 points with  
$\log \Delta F <0.6$  we can obtain   
a moderate correlation up to r$_P \sim 0.4$).  
  
During the time  of these observations MWC~560 is in the process of   
jet ejection, and the jet is even precessing (Iijima 2002).  
The lack of significant correlation between the flickering amplitude and   
hot component flux probably is a result of the outflow,  
i.e. the jet ejection destroys the  innermost parts   
of the accretion disk, where the flickering is formed.  
  
The connection between the jet and flickering is not investigated in MWC~560,  
however it is visible in the 1997 jet launch in CH~Cyg  
(Sokoloski \& Kenyon 2003a).  
The disks and jets are also connected in quasars and microquasars  
(see  Livio, Pringle \& King 2003 and references therein).  
In this sense the fact that the correlation between  
$\Delta$ F and F$_{av}$ is loose in MWC~560, is probably due to the  
outflow and its connection  with the accretion disk.  
  
\section{Discussion of flickering amplitude}  
  
In all three symbiotic stars with available data the flickering amplitude   
shows the same tendency to increase with increases of the hot component  
flux. If we accept that there are no different states in CH~Cyg and   
all points lay on the same line it means that the obtained value of   
the slope is very similar to T~CrB, which in terms of the Bruch \& Dushl (1993) model  
in turn means that in both stars the size of the boundary layer remains constant  
(see Sect.\ref{BL}). In MWC~560   
the flickering is (probably) influenced by the outflow and $\Delta$F depends  
weakly on F$_{av}$, and the correlation is not well defined.   
  
The other possibility that the flickering amplitude of CH~Cyg lays on two parallel   
lines would give different values of k in the relationship   
$\Delta F \propto (F)^k$.  
One of the reasons for this difference could be the magnetic field   
of the WD. Here we want to point out that the flickering amplitude   
could be connected with the magnetic field.   
The most probable place for the origin of the flickering  
is the inner parts of the accretion disk. If the flickering is a  
result of the turbulence in the inner parts of the disk  
then the energy available in the turbulence will be proportional  
to the density where the flickering forms (Bruch, 1992).  
If the white dwarf is magnetic, the inner parts of the  
accretion disk will be destroyed by the magnetic field.  
Different instabilities can appear then at the inner edge of the disk.  
These instabilities permit the accreting material to be absorbed  
from the magnetosphere as blobs.  The energy releasing will  
be unsteady and we suppose that the amplitude of the flickering  
will be proportional to the typical mass of the blobs, and the mass  
of the blobs will in turn be proportional to the density  
at the inner edge of the disk.  The density in the disk  
can be estimated as (Lipunov 1992):  
  
\begin{equation}  
 \rho_{in} \approx  \alpha^{-1} \left( \frac{R}{H} \right)  
        \frac {\dot {M_a}} { 4 \pi R^{3/2} \sqrt {2GM}  },  
  \label{eq2}%
\end{equation}  
where $(R/H)$ is the ratio between the radius and the vertical size  
of the disk, ($R/H$) is usually adopted to be a constant of order 0.01--0.1.  
$M$ is the mass of the white dwarf.  
  
  If the white dwarf is non-magnetic, the inner radius  of the disk will be  
approximately equal to the white dwarf radius (for a thin boundary layer) and  
consequently the density at the inner edge is given by $\rho_{in} \propto \dot M_a$.  
The same relationship will be fulfilled if the boundary layer is not thin and its  
size does not change.  
If the white dwarf is magnetic the radius $R_0$ of the inner disk edge  
may be expressed as (Lamb, Pethick \& Pines 1973):  
   \begin{equation}  
 R_0=N (GM)^{-1/7} \mu ^{4/7} \dot {M_a}^{-2/7}  ,  
   \label{eq3}%
   \end{equation}  
where $N$ is a constant of order 1, and $\mu$ is the white dwarf magnetic moment.  
In this case (from  Eqn. \ref{eq2} and \ref{eq3})  
   \begin{equation}  
\rho_{in} \propto \dot M_a^{k},\ \   k = \frac {10}{7}  
   \label{eq4}%
   \end{equation}  
where $k = \frac {10}{7}=1.43 $  is in agreement with  
the behaviour of CH~Cyg   
(Miko{\l}ajewski et al., 1990, see also Sect.\ref{amplCH} ),  
if the suppositions (1) $\Delta F \propto \rho_{in}$,   
and (2) accretor-propeller states in CH~Cyg (Miko{\l}ajewski et al. 1990)   
are correct.  
  
The data for T~CrB are consistent with $k=1$ as expected   
for a low or non-magnetic white dwarf, i.e. the position of the inner edge   
of the accretion disk  does not depend on the mass accretion rate.   
  
The presence of a magnetic WD in CH~Cyg is not a certain fact.   
Sokoloski \& Kenyon (2002a), Crocker et al. (2001), Ezuka et al. (1998)  
threw doubts about the presence of such a magnetic WD in CH~Cyg.  
However, the magnetic propeller model of   
(Miko{\l}ajewski \& Miko{\l}ajewska, 1988) still 
remains the most promising for the variability of this object.   
  
If the differences in the behaviour of the flickering   
in T~CrB, CH~Cyg, and MWC~560 are not connected with the magnetic field   
and jet ejection, other possible reasons   
may be the changes of the energy distribution, or different   
mechanisms generating the flickering in these objects.

\section{Conclusions}  
  
We have analysed the U band variability of the recurrent nova   
and symbiotic star T~CrB, and compared its behaviour with two   
other symbiotic stars  CH~Cyg and MWC~560.   
During the period of our observations T~CrB brightness 
varies in between U$=$13 and U$=$10 mag.   
The analyses we performed show that:  
(1) the V brightness during the last 22 years is dominated by the ellipsoidal variability  
of the red giant, however the hot component variability with  
$\Delta U \simgt $2 mag,  
introduces a shift in V with about 0.15 mag.  
No signs of variability of the red giant has been detected.  
(2) the power spectrum of the flickering does not change during our observations,  
remaining always with slope $\gamma \approx -1.5$  in spite of   
the changes in U. We do not detect changes in the power  
spectrum like those observed in CH~Cyg;  
(3) The calculated e-folding time of the ACF also does not   
show dependence on the changes in U;   
(4) The flickering amplitude is strongly correlated with the   
average flux of the hot component.   
(5) The differences in the dependence of the flickering amplitude   
between T~CrB, CH~Cyg and MWC~560  
could be connected with jet ejection and the possible presence   
of a magnetic white dwarf in the last two.  
In general, in T~CrB, we have observed  flickering, which does not
change considerably its characteristics (at least during the time of our observations).
  
In the future it would be very interesting to determine the behaviour of   
the flickering amplitude, ACF, power spectra, etc.  
of other symbiotic stars with flickering (in particular RS Oph, RT Cru, $o$ Ceti)   
as well as the flickering  of MWC~560 during phases without outflow,   
as well the connection of flickering with jet precession.  
Simultaneous spectral and photometric observations over a wide spectral range  
from UV to IR  could be very useful  to investigate in detail  
the flickering behaviour and its connection  
with accretion disk instabilities and jet ejections.    
  
\section*{Acknowledgments}  
  
This research has made use of SIMBAD, IRAF, and  Starlink.  
RZ is supported by a PPARC Research Assistantship  
and MFB is a PPARC Senior Fellow.

\end{document}